# Absence of Superconductivity in $Nd_{0.8}Sr_{0.2}NiO_x$ Thin Films without Chemical Reduction

Xiao-Rong Zhou, Ze-Xin Feng, Pei-Xin Qin, Han Yan, Shuai Hu, Hui-Xin Guo, Xiao-Ning Wang, Hao-Jiang Wu, Xin Zhang, Hong-Yu Chen, Xue-Peng Qiu, Zhi-Qi Liu *(iD 0000-0001-7538-4879)

**Abstract**  The recently reported superconductivity 9-15 K in $Nd_{0.8}Sr_{0.2}NiO_2$/$SrTiO_3$ heterostructures that were fabricated by a soft-chemical topotactic reduction approach based on precursor $Nd_{0.8}Sr_{0.2}NiO_3$ thin films deposited on $SrTiO_3$ substrates, has excited an immediate surge of research interest. To explore an alternative physical path instead of chemical reduction to realizing superconductivity in this compound, using pulsed laser deposition, we systematically fabricated 63 $Nd_{0.8}Sr_{0.2}NiO_x$ (NSNO) thin films at a wide range of oxygen partial pressures on various oxide substrates. Transport measurements did not find any signature of superconductivity in all the 63 thin-film samples. With the oxygen content reducing in the NSNO films by lowering the deposition oxygen pressure, the NSNO films are getting more resistive and finally become insulating. Furthermore, we tried to cap a 20-nm-thick amorphous $LaAlO_3$ layer on a $Nd_{0.8}Sr_{0.2}NiO_3$ thin film deposited at a high oxygen pressure of 20 Pa to create oxygen vacancies on its surface and did not succeed in obtaining higher conductivity either. Our experimental results together with the recent report on the absence of superconductivity in synthesized bulk $Nd_{0.8}Sr_{0.2}NiO_2$ crystals suggest that the chemical reduction approach could be unique for yielding superconductivity in NSNO/$SrTiO_3$ heterostructures. However, $SrTiO_3$ substrates could be reduced to generate oxygen vacancies during the chemical reduction process as well, which may thus partially contribute to conductivity.

X.-R. Zhou, Z.-X. Feng, P.-X. Qin, H. Yan, H.-X. Guo, X.-N. Wang, H.-J. Wu, X. Zhang, H.-Y. Chen, Z.-Q. Liu*
School of Materials Science and Engineering, Beihang University, Beijing 100191, China
e-mail: zhiqi@buaa.edu.cn

S. Hu, X.-P. Qiu
Shanghai Key Laboratory of Special Artificial Microstructure Materials and Technology, and Pohl Institute of Solid-State Physics, School of Physics Science and Engineering, Tongji University, Shanghai 200092, China

# 1 Introduction

In contrast to the antiferromagnetic ordering of cuprates that has been generally believed to be a key factor for high-temperature superconductivity [1], the parent compound $NdNiO_2$ of the recently reported superconducting $Nd_{0.8}Sr_{0.2}NiO_2$ thin films lacks in any magnetic ordering down to 1.7 K [2]. Therefore, the recent report of 9-15 K superconductivity in $Nd_{0.8}Sr_{0.2}NiO_2$/$SrTiO_3$ (STO) heterostructures [3] may represent an important new type of superconductivity, despite of the similar electronic structure of $Cu^{2+}$ and $Ni^+$.

The superconductivity emerged after the soft-chemical topotactic reduction approach, which was demonstrated to reduce the non-superconducting $Nd_{0.8}Sr_{0.2}NiO_3$/STO heterostructures into superconducting $Nd_{0.8}Sr_{0.2}NiO_2$/STO heterostructures. Nevertheless, during this chemical reduction approach, the STO single-crystal substrates can be reduced to generate oxygen vacancies as well. As the defect levels of oxygen vacancies in STO are shallow, ~4-25 meV below the bottom of the conduction band, they can serve as shallow electron donors to make STO itself metallic [4, 5]. In addition, it was emphasized that the polar discontinuity exists between nonpolar $Nd_{0.8}Sr_{0.2}NiO_2$ films and STO substrates which may contribute to interfacial conductivity and the epitaxial strain effect could be helpful for stabilizing the superconducting $Nd_{0.8}Sr_{0.2}NiO_2$ phase [3], it is of immediate interest to find an alternative non-chemical reduction approach for varying the oxygen content of $Nd_{0.8}Sr_{0.2}NiO_x$ (NSNO) films and further examine the transport properties of NSNO films grown onto various substrates with different lattice constants and polar natures.

Unlike the high-temperature superconductivity in cuprates [6] and Fe-based superconductors [7], which was first discovered in bulk compounds, this Ni-based superconductivity has been discovered in thin-film samples [3], which inevitably include substrate and heterointerface effects. Especially, the recent transport study on bulk $Nd_{0.8}Sr_{0.2}NiO_2$ crystal samples did not find any signature of superconductivity or even metallicity in bulk $Nd_{0.8}Sr_{0.2}NiO_2$ crystals under both zero and a 50.2 GPa pressure [8], which thus has called in question the intrinsic superconductivity in this Ni-based oxide and has further emphasized the important roles of film/substrate interface effects in thin film heterostructure samples as the origin of the exotic superconductivity. In this work, we tried to examine the feasibility of the physical approach to tuning the oxygen content in the NSNO thin films by varying the deposition oxygen pressure in pulsed laser deposition for realizing possible superconductivity in this compound, which is expected to help clarify the material source of the reported superconductivity.

## 2 Experimental

$Nd_{0.8}Sr_{0.2}NiO_x$ films were grown by using pulsed laser deposition with a 248-nm excimer laser. The substrate temperature was kept at 600 °C during deposition. The laser fluence was 1.6 J·cm$^{-2}$ and the repetition rate was 10 Hz. A polycrystalline NSNO ceramic target was sintered at 1500 °C under vacuum to produce a possible $Nd_{0.8}Sr_{0.2}NiO_2$ phase. The target-substrate distance was fixed at 60 mm. The ramp rate for both heating and cooling was 20 °C·min$^{-1}$ and the cooling was accompanied by the deposition oxygen partial pressure. Electrical measurements were conducted in a Quantum Design physical property measurement system. Electrical connections of the samples were prepared using aluminum wire with a diameter of 30 μm by wire bonding. The structure of thin film samples were characterized by a four-circle Bruker D8 Discover X-ray diffractometer and a JEOL JEM-2200 FS transmission electron microscopy setup.

## 3 Results and discussion

In this work, by using pulsed laser deposition with a 248-nm excimer laser, we first tried to modulate the oxygen content in NSNO films by altering the deposition oxygen partial pressure from 20 (the same deposition oxygen pressure as in Ref.[3]), $1.33\times10^1$, $1.33\times10^0$, $1.33\times10^{-1}$, $1.33\times10^{-2}$, $1.33\times10^{-3}$ to $1.33\times10^{-4}$ Pa. The thin film thickness was fixed at 70 nm, which was calibrated by X-ray reflectivity measurements. For each oxygen growth partial pressure, 9 types of single-crystal oxide substrates were utilized, which were (110)-oriented $YAlO_3$ (YAO), (001)-oriented $LaAlO_3$ (LAO), (001)-oriented $(LaAlO_3)_{0.3}(Sr_2AlTaO_6)_{0.7}$ (LSAT), (110)-oriented $NdGaO_3$ (NGO), (001)-oriented $SrTiO_3$ (STO), (110)-oriented $DyScO_3$ (DSO), (001)-oriented $0.7PbMg_{1/3}Nb_{2/3}O_3$-$0.3PbTiO_3$ (PMN-PT), (001)-oriented $MgAl_2O_4$ (MAO), and (001)-oriented MgO. Their equivalent in-plane lattice constants are schematized in Fig. 1a.

XRD pattern of the polycrystalline NSNO ceramic target sintered at 1500 °C under vacuum is shown in Fig. 1b, which reveals a mixture of the $Nd_{0.8}Sr_{0.2}NiO_2$ and $Nd_{0.8}Sr_{0.2}NiO_3$ phases [8]. The temperature-dependent resistivity of the polycrystalline target measured via the linear four-probe method is plotted in Fig. 1c. It exhibits insulating behavior with room-temperature resistivity of ~0.48 Ω·cm, and is in good agreement with the transport properties and the absence of superconductivity in synthesized bulk $Nd_{0.8}Sr_{0.2}NiO_2$ crystals [8].

Figure 2a plots XRD spectra of the NSNO films deposited on STO at different oxygen pressures. The epitaxial $Nd_{0.8}Sr_{0.2}NiO_3$ phase is clearly formed for high oxygen pressures such as 20 and $1.33\times10^1$ Pa, which is consistent with the report in Ref. [3]. However, with the oxygen partial pressure and the oxygen content reducing in NSNO films, the $Nd_{0.8}Sr_{0.2}NiO_3$ (002) peak disappears starting from $1.33\times10^0$ Pa. As the crystallization temperature for most of complex oxides is ~450 °C [9] and all our depositions have been performed at a substrate temperature of 600 °C, the suppression of the $Nd_{0.8}Sr_{0.2}NiO_3$ (002) peak in the NSNO films deposited at low oxygen pressures does not indicate that the NSNO films become amorphous, but may imply the formation of another NSNO phase with less oxygen. The cross-sectional TEM image of a NSNO/STO heterostructure fabricated at $1.33\times10^{-4}$ Pa is shown in Fig. 2b, which indicates an ~6-nm-thick amorphous interfacial layer and a nano-crystalline layer of the $Nd_{0.8}Sr_{0.2}NiO_x$ film.

The temperature-dependent sheet resistance ($R_S$-$T$) for the NSNO films deposited at 20 Pa oxygen pressure onto various substrates are summarized in Fig. 3a. Overall, there is a remarkable substrate effect on the transport properties of NSNO films. The sheet resistance ($R_S$) at 300 K is varied by more than two orders of magnitude from LAO to MgO. The NSNO films grown on oxide substrates with relatively small in-plane lattice constants including YAO, LAO, LSAT, NGO, STO and DSO exhibit metallic behavior at high temperatures and a small resistance upturn at ~50, ~10, ~20, ~20, ~65 and ~50 K, respectively.

For substrates with large in-plane lattice constants such as PMN-PT, MAO and MgO, the thin films show insulating behavior from room temperature. As bulk NSNO has in-plane lattice constants smaller than 0.4 nm, for example, its $Nd_{0.8}Sr_{0.2}NiO_3$ phase has a pseudocubic lattice structure with $a \approx 0.381$ nm and the $Nd_{0.8}Sr_{0.2}NiO_2$ phase exhibits a tetragonal structure with $a \approx 0.392$ nm [3], NSNO films feel the tensile strain from PMN-PT, MAO and MgO substrates. Consequently, these transport measurements suggest that the electronic state of NSNO films is rather sensitive to substrates and a tensile strain can lead to an insulating state.

Regarding the polar discontinuity effect [10-12] at the interface between polar NSNO films and oxide substrates on the electrical conductivity of the NSNO films as emphasized in Ref. [3], the NSNO/LAO and NSNO/STO heterostructures may be good examples for comparison. As LAO is a polar oxide and STO is a non-polar oxide, there is no polar discontinuity at the interface between the NSNO film and the LAO substrate and thus there shall be no electronic reconstruction within the NSNO films [12] or across the interfaces [4, 11]. However, the NSNO/LAO heterostructure exhibits much smaller sheet resistance compared with the NSNO/STO heterostructure. This indicates that the polar discontinuity effect is not a predominant effect in determining the electrical conductivity of NSSO thin films.

The $R_S$-$T$ of NSNO films deposited at $1.33\times10^1$ Pa oxygen pressure (Fig. 3b) generally have similar substrate dependence, i.e., they are insulating while growing on PMN-PT, MAO and MgO substrates but are metallic at high temperature for other substrates. The low-temperature resistance upturn occurs at ~113, ~30, ~52, ~52, ~100, ~162 K for YAO, LAO, LSAT, NGO, STO and DSO single-crystal oxide substrates, respectively. The obviously enhanced resistance upturn temperatures indicate that the NSNO films are more insulating while reducing the oxygen content.

Compared with the NSNO films fabricated at 20 and $1.33\times10^1$ Pa, the NSNO films deposited at 1.33 Pa (Fig. 4a) and $1.33\times10^{-1}$ Pa (Fig. 4b) possess ten times larger sheet resistance and are all insulating from room temperature. Furthermore, the NSNO films deposited at $1.33\times10^{-2}$ Pa are even more resistive (Fig. 4c). Finally, all the NSNO films deposited at $1.33\times10^{-3}$ and $1.33\times10^{-4}$ Pa are highly insulating, and their room-temperature sheet resistance is beyond our measurement limit of ~109 $\Omega\cdot\square^{-1}$ (Fig. 4d).

Up to now, our endeavor on trying to reduce the NSNO films by lowering deposition oxygen pressure to induce possible superconductivity has failed. On the other hand, the other effective way to reducing oxygen at oxide surfaces can be the room-temperature deposition of an additional LAO layer in a vacuum condition such as $1.33\times10^{-4}$ Pa by pulsed laser deposition [4, 13]. That is because the Al atoms excited by pulsed laser from the LAO target are of superior chemical affinity to oxygen atoms, and thus they can actively capture oxygen atoms from oxide substrate surfaces to form the stable LAO phase, consequently reducing the surface oxygen content in substrates.

We first chose the NSNO/LAO heterostructure that is fabricated at 0 Pa and exhibits the lowest resistance, and then deposited a 20-nm-thick amorphous LAO (aLAO) layer on top of it under a vacuum condition of $1.33\times10^{-4}$ Pa at room temperature to form an aLAO/NSNO/LAO heterostructure. After deposition, as opposite to our motivation for achieving better conductivity, the sheet resistance of the heterostructure is instead increased by ~50% at room temperature (Fig. 5). Nevertheless, this is consistent with our oxygen partial pressure modulation experiments, i.e., the NSNO films are not getting more metallic or superconducting while reducing the oxygen content, but become more resistive.

In addition, the magnetic properties of our NSNO films fabricated at different oxygen pressures could be interesting as they may lead to exotic spintronic phenomena [14] and more importantly may provide additional evidence for revealing the possibility of superconductivity. Especially, if the antiferromagnetic order could still exist in these thin-film samples, the perspective of this field would be substantially changed. On the other hand, as antiferromagnetism is getting more interesting due to the recent rapid development of antiferromagnetic spintronics such as in some recent reports related to our own studies [15-31], the effective control of the possible magnetic order in this compound could further create a new path to tuning superconductivity.

## 4 Conclusion

The vacuum-sintered polycrystalline target has a mixture of the $Nd_{0.8}Sr_{0.2}NiO_2$ and $Nd_{0.8}Sr_{0.2}NiO_3$ phases and shows insulating transport behavior. The transport properties of NSNO films are strongly dependent on substrates and a tensile strain leads to a more resistive state. In sharp contrast to the superconductivity discovered in the $Nd_{0.8}Sr_{0.2}NiO_2$ films obtained by chemical reduction, NSNO thin films fabricated onto different substrates at different oxygen pressures are getting more and more insulating by reducing the oxygen content, which is consistent with the recently reported insulating transport behavior in bulk $Nd_{0.8}Sr_{0.2}NiO_2$ compounds. These results suggest that the superconductivity may not be an intrinsic property of the $Nd_{0.8}Sr_{0.2}NiO_2$ compound. Instead, the chemical reduction process may have induced other effects in thin films or even in the STO substrates in addition to the reduction of the oxygen content of NSNO films from $Nd_{0.8}Sr_{0.2}NiO_3$ to $Nd_{0.8}Sr_{0.2}NiO_2$, which needs to be further figured out. Overall, the convincing and intrinsic superconductivity in a Ni-based oxide may remain yet to be demonstrated.

**Acknowledgments** This work was financially supported from the National Natural Science Foundation of China (Nos. 51822101, 51861135104, 51771009 and 11704018).

Figures

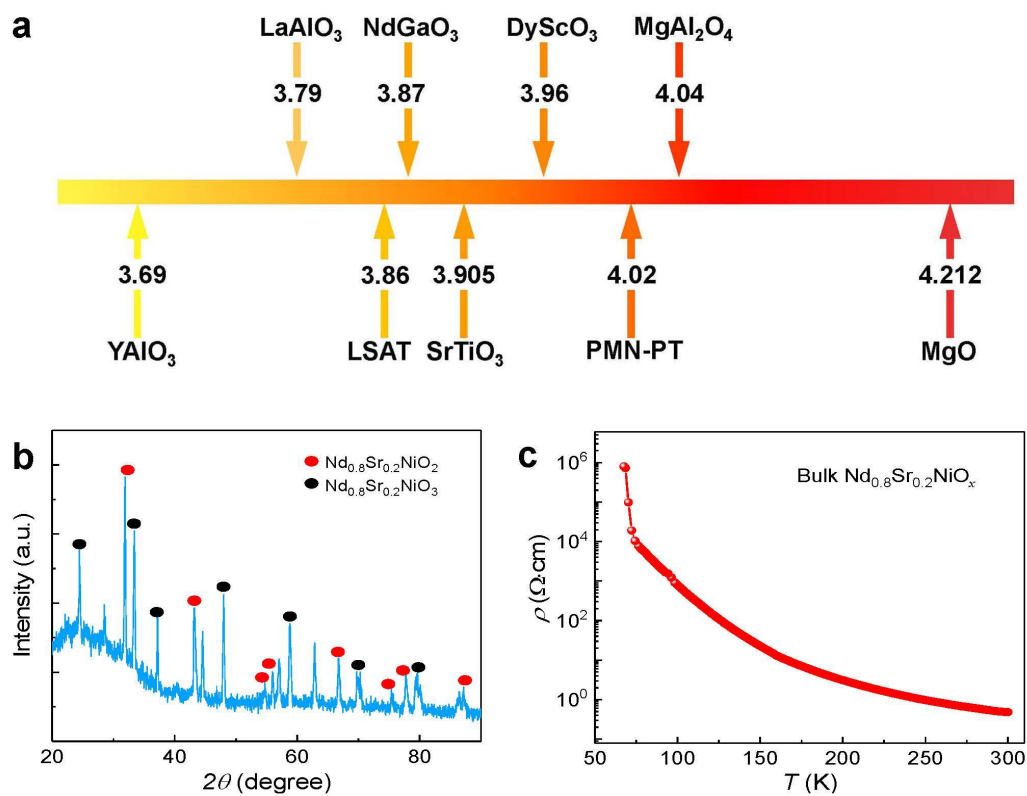

Fig.1 The substrates and the vacuum-sintered $Nd_{0.8}Sr_{0.2}NiO_x$ (NSNO) target. **a** Equivalent in-plane lattice constants for pseudocubic or cubic perovskite oxide substrates. From left to right: (110)-oriented $YAlO_3$, (001)-oriented $LaAlO_3$, (001)-oriented LSAT, (110)-oriented $NdGaO_3$, (001)-oriented $SrTiO_3$, (110)-oriented $DyScO_3$, (001)-oriented PMN-PT, (001)-oriented $MgAl_2O_4$ and (001)-oriented MgO. **b** Powder X-ray diffraction spectrum of the vacuum-sintered $Nd_{0.8}Sr_{0.2}NiO_x$ (NSNO) target. Red solid circles mark the peaks of the $Nd_{0.8}Sr_{0.2}NiO_2$ phase and black solid circles indicate the diffraction peaks of the $Nd_{0.8}Sr_{0.2}NiO_3$ phase [3]. **c** Temperature-dependent resistivity of the bulk NSNO target.

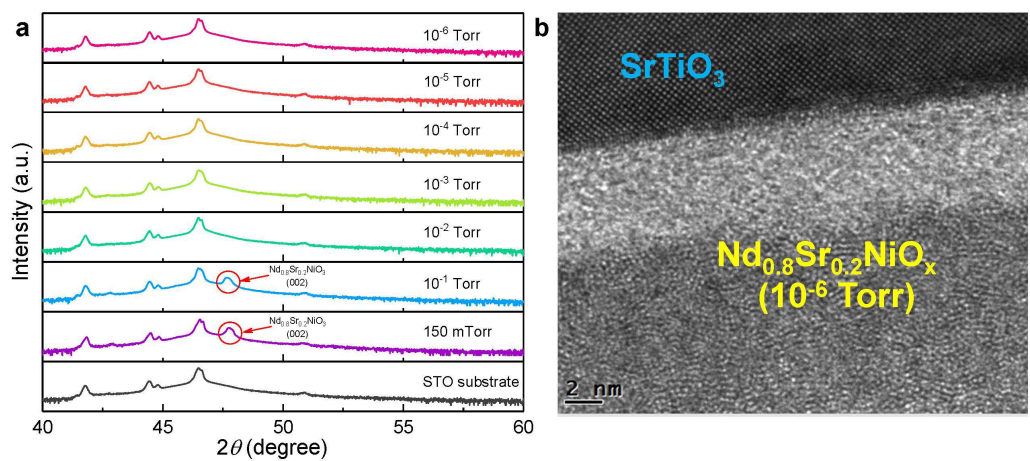

Fig.2 **a** X-ray spectra of NSNO films grown on $SrTiO_3$ (STO) at different oxygen pressures ranging from 150 mTorr to $10^{-6}$ Torr. The X-ray source includes multiple wavelengths and thus there are multiple diffraction peaks for the (002) diffraction of a STO single-crystal substrate. **b** Cross-section transmission electron microscopy image of a NSTO/STO heterostructure deposited at $10^{-6}$ Torr, indicating an ~6-nm-thick amorphous interface layer and an nano-crystalline layer of the $Nd_{0.8}Sr_{0.2}NiO_x$ film.

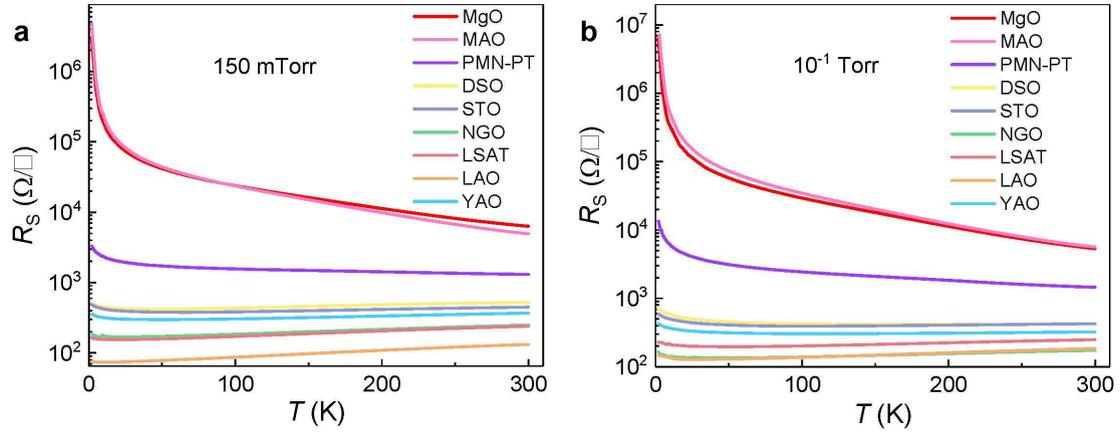

Fig.3 $R_S$-$T$ curves for NSNO films deposited on various oxide substrate at 150 mTorr and $10^{-1}$ Torr. **a** Temperature-dependent sheet resistance ($R_S$-$T$) for NSNO films deposited on various oxide substrates at 150 mTorr oxygen pressure. **b** $R_S$-$T$ curves for NSNO films deposited on various oxide substrates at $10^{-1}$ Torr oxygen pressure.

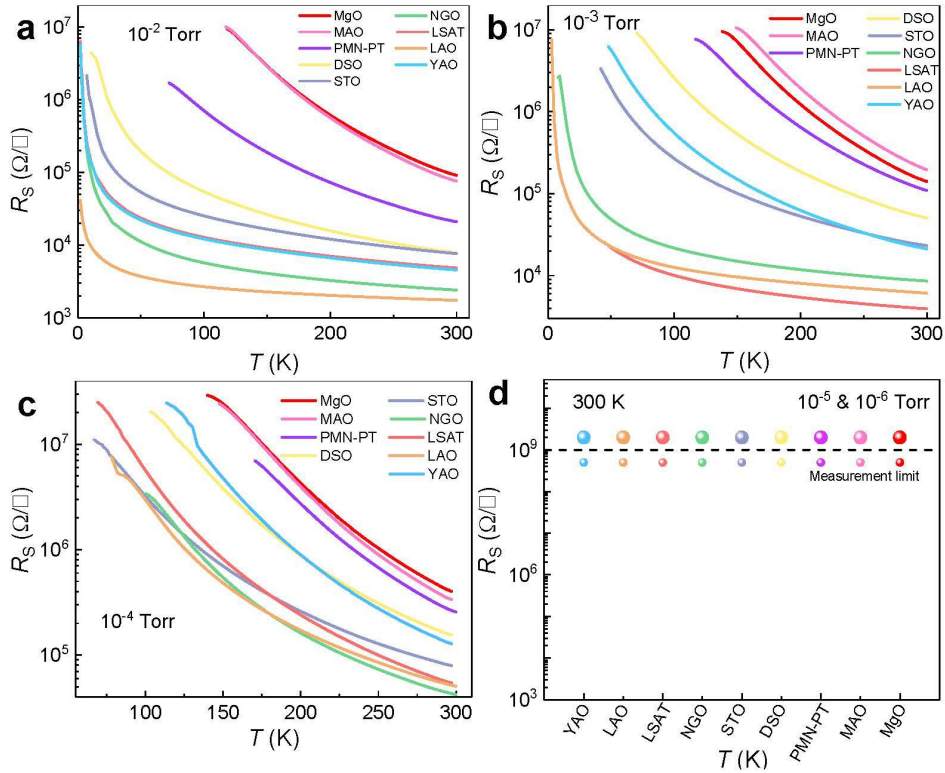

Fig.4 $R_S$-$T$ curves for NSNO films deposited on various oxide substrate at $10^{-2}$, $10^{-3}$, $10^{-4}$, $10^{-5}$ & $10^{-6}$ Torr. **a** RS-T curves for NSNO films deposited on various oxide substrates at $10^{-2}$ Torr oxygen pressure. **b** $R_S$-$T$ curves for NSNO films deposited on various oxide substrates at $10^{-3}$ Torr oxygen pressure. **c** $R_S$-$T$ curves for NSNO films deposited on various oxide substrates at $10^{-4}$ Torr oxygen pressure. **d** Room-temperature sheet resistance of NSNO films deposition onto various oxide substrates at $10^{-5}$ Torr (smaller solid balls) and $10^{-6}$ Torr (larger solid balls) oxygen pressures.

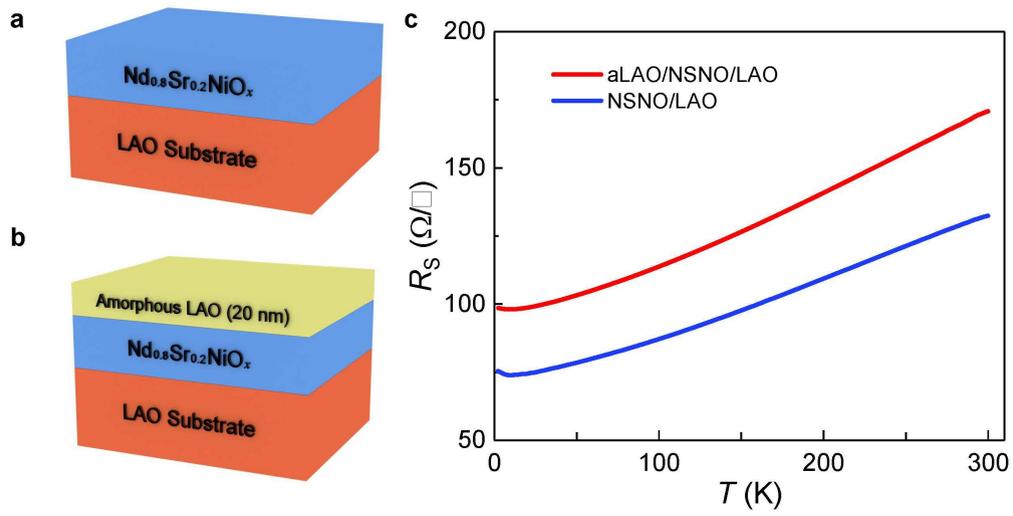

Fig.5 Amorphous LaAlO$_3$ (LAO) capping experiment. **a** & **b** Schematic of the NSNO/LAO and aLAO/NSNO/LAO heterostructure, respectively. **c** $R_S$-$T$ curves for the two types of heterostructures.